\documentclass[12pt,amsmath,floats,floatfix,superscriptaddress]{article}
\usepackage{graphicx} \usepackage{bm} \usepackage{epsfig}
\usepackage{amsmath, amssymb}
\usepackage{wrapfig}
\usepackage{paralist}
   \usepackage{cite}
\usepackage{setspace}
\usepackage{color}
\usepackage{caption}
\usepackage{colordvi}
\usepackage{comment}
\usepackage[export]{adjustbox}
\usepackage{endnotes}

\newcommand{\beq}{\begin{equation}}
\newcommand{\eeq}{\end{equation}}
\newcommand{\bea}{\begin{eqnarray}}
\newcommand{\eea}{\end{eqnarray}}
\newcommand{\bit}{\begin{itemize}\setlength\itemsep{0em}}
\newcommand{\eit}{\end{itemize}}
\newcommand{\im}{\item}

\begin{document}
\begin{center}
{\fontsize{20}{28}\selectfont  \sffamily  The Music of the Spheres:\\[0.2cm] The Dawn of Gravitational Wave Science}
\\[0.5cm]
{\fontsize{12}{16}\selectfont \sffamily Rafael A. Porto}
\\[0.2cm]
 { \it ICTP South American Institute for Fundamental Research}
 \\ Rua Dr. Bento Teobaldo Ferraz 271, 01140-070 S\~ao Paulo, SP Brazil\vskip 8pt
 
 \end{center}



\vskip 1cm

Black holes --the end product of gravitational collapse-- span a vast range of scales, from Planck 
size\footnote{The Planck length is given by $\ell_{\rm P}  = \sqrt{\hbar G_N/c^3} \sim 10^{-35}$\,m. With $(G_N,\hbar,c)$ the Newton and Planck constants and the speed of light, respectively.} to ultra-massive ones, as heavy as billions of solar masses (whose horizons are larger than the distance between the sun and the earth, the astronomical unit), which are expected to be contained in quasars.
\begin{center}
\vspace{-0.1cm}
\includegraphics[width=0.75\textwidth]{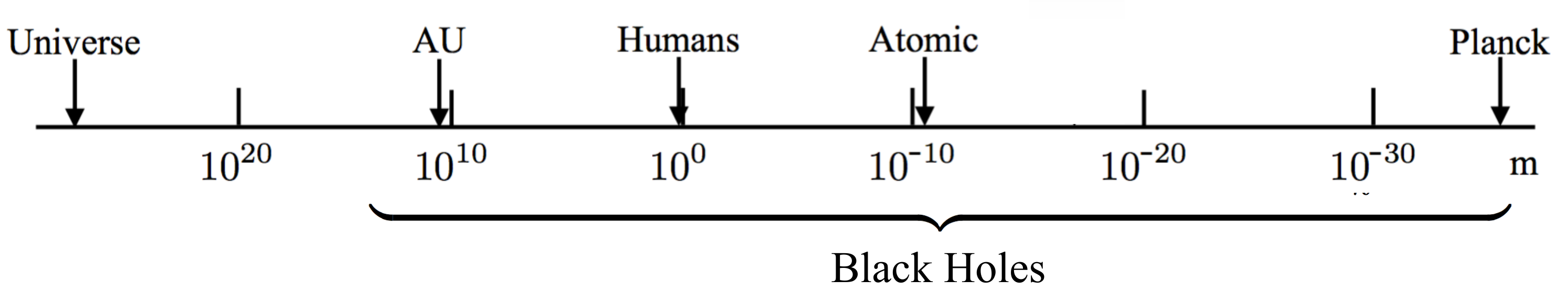}
\vspace{-0.2cm}
\end{center}
They are some of the most awe-inspiring objects in nature, yet Einstein himself believed to never form in physical processes. In~isolation, and in the classical realm, non-rotating electrically neutral black holes are described by the Schwarszchild-Droste solution to general relativity, Einstein's theory of gravity, found readily after the equations were presented a hundred years ago. For spinning black holes, on the other hand, we have the Kerr solution, discovered after Einstein's death. This illustrates how difficult the non-linear equations of general relativity really are, since simply incorporating rotation took almost fifty years!\vskip 4pt Black holes feature physical (and unphysical, depending on the choice of coordinates) singularities, spacetime regions from which light cannot escape (horizons),  regions from which energy can be extracted (ergospheres), and even temperature in a world with a non-zero Planck constant. In company, for instance in a binary system,  exact solutions are not known analytically, except in very especial configurations. This is due to the proliferation of different relevant scales (sizes, separation, velocities and typical wavelength of the emitted radiation) and the non-linearities of Einstein's~gravity. Nonetheless, one can solve Einstein's equations perturbatively during the so-called {\it inspiral} phase, where the black hole's separation is large enough (suffices a hundred times their size apart) that, even though they may be orbiting each other much (much) faster than we are accustomed to  travel, they do so at a fraction of the speed of light. This is the regime a typical binary system will spend most of its lifetime dwelling upon. The perturbative expansion is ultimately characterized in powers of the relative velocity over the speed of light, $v/c$, and it is commonly referred as the `Post-Newtonian' approximation.\vskip 4pt
\begin{wrapfigure}{r}{0.35\textwidth}
\vspace{-0.4cm}
\begin{center}
\includegraphics[width=0.35\textwidth]{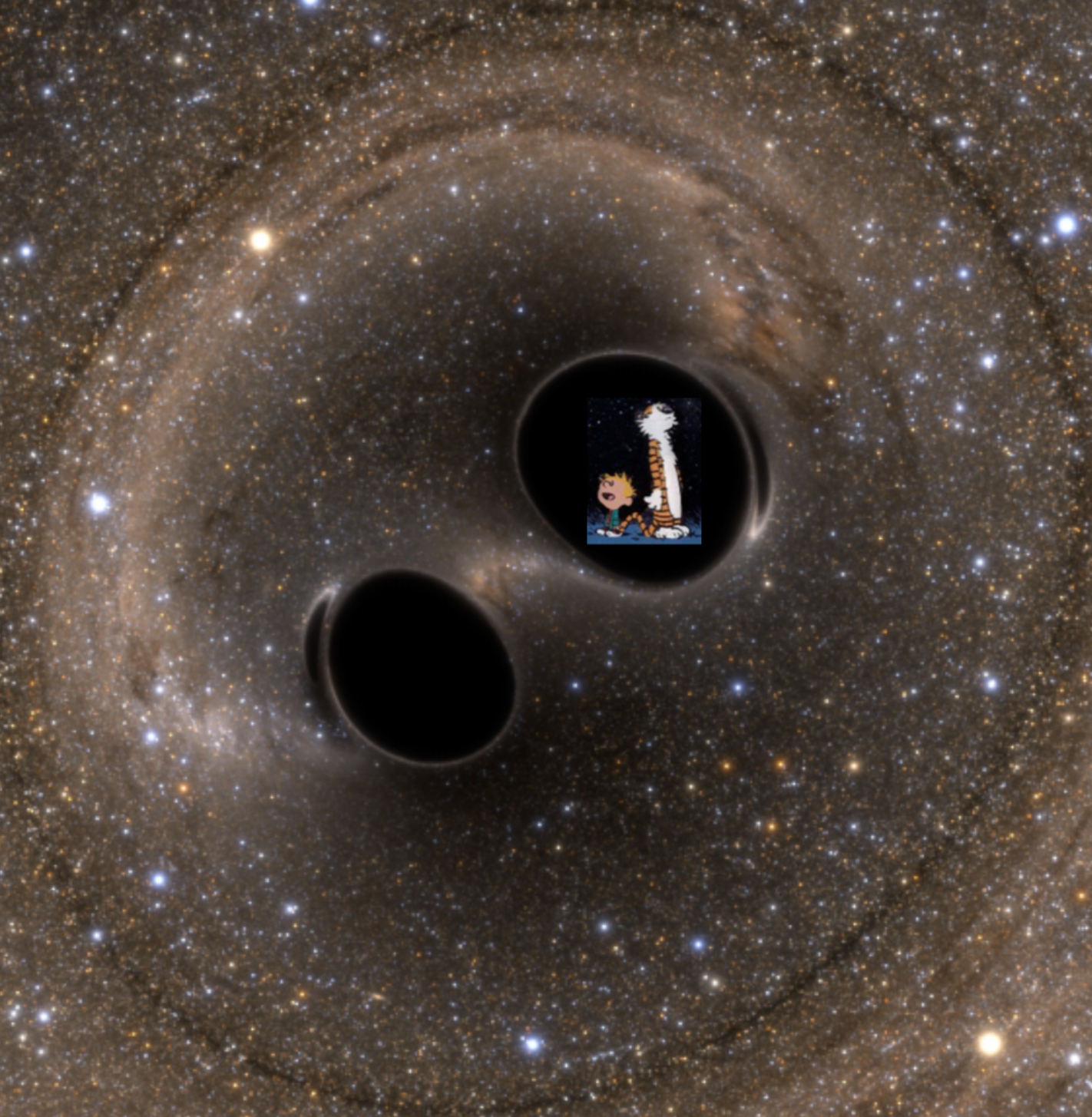}
\end{center}
\vspace{-0.7cm}
\end{wrapfigure}
The two-body problem in gravity has been the object of the greatest interest for centuries. Since the time of Galileo and Newton (and Hooke and Halley and others), we know planets move in ellipses according to an inverse square law. At its full expression, the binary problem in general relativity, however, is in sharp contrast to Newtonian gravity. The latter is a linear theory which does not feature a limit for the speed of propagation, hence lacking gravitational radiation, and therefore --up to tidal effects-- can be solved exactly even for comparable~masses. In Einstein's theory on the other hand, we not only have non-linear interactions but we also encounter dissipation (and absorption) in the form of gravitational waves. Similarly to dipole radiation in electromagnetism, the celebrated quadrupole formula\footnote{Einstein produced two papers on approximate wave-like solutions to general relativity. Not only his first attempt was incorrect, his second version was off by a factor of 2.} gives us a quantifiable expression for the total power emitted from a binary system. The rate has been tested with remarkable precision observing the orbital decay of the so-called `Hulse-Taylor binary pulsar', or PSR 1913+16 for short. Gravitational waves, as ripples of spacetime, carry energy away from the  system precisely as predicted by general relativity, and a Nobel prize for this discovery was already awarded. PSR 1913+16 will not merge anytime soon, and therefore it provides a {\it weak field} test of Einstein's theory. However, it already evidenced the reality of gravitational wave emission in nature, settling a long-lasting debate about their own existence which goes back to Einstein himself. Later in life, and together with a young collaborator, Einstein claimed to have proven that his full theory did not allow for exact gravitational wave solutions. This paper was submitted to the `Physical Review' journal and, as it is normal practice in today's academic world, it was first sent for peer review to a another physicist. In this case, the renowned cosmologist Howard P. Robertson, without the author's knowledge. Robertson pointed out a mistake and suggested some revisions. This~prompted Einstein's response: `We~had sent you the paper for publication and had not authorized you to show it to specialists before it is printed. I see~no~reason to address the --in any case erroneous-- comments of your anonymous expert.' The final version --this time without the original claims (nor a reference to the previous interaction)-- eventually appeared in print in another journal.\vskip 4pt 
\begin{wrapfigure}{r}{0.17\textwidth}
\vspace{-0.9cm}
\begin{center}
\includegraphics[width=0.17\textwidth]{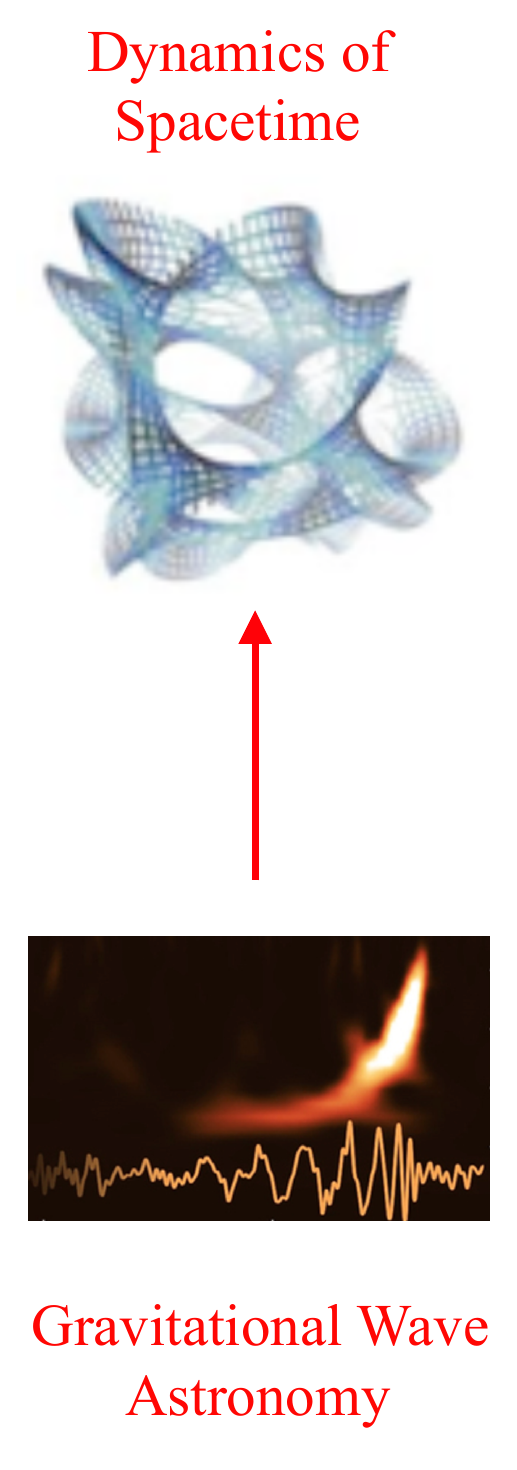}
\end{center}
\vspace{-0.7cm}
\end{wrapfigure}
Because gravity is so feeble, it wasn't until the advent of earth-based gravitational wave observatories such as LIGO that the direct observation of the ripples of spacetime was made possible. Almost one hundred years after general relativity was born, on September 14th of 2015, the LIGO scientific collaboration detected a transient gravitational wave signal produced --about a billion years ago-- by a black hole binary system of approximately sixty five solar masses in total, prior to impact.\footnote{ LIGO, the Laser Interferometer Gravitational-Wave Observatory, is able to measure changes in the arm's length with a precision that is equivalent to determining the distance to the star closest to the solar system, Alpha Centauri, with the accuracy of a human~hair!} This cataclysmic event, which merely induced a tiny wiggle on LIGO's detectors, shed about three solar masses in energy (recall  the most famous equation in physics: $E=mc^2$) in less than a second.\footnote{The emitted power, at its peak, was greater than the one obtained by combining all luminous matter in the observable universe. For comparison, the sun-earth binary system emits in gravitational waves the equivalent to a light bulb of roughly 200 Watts. (Thankfully, we receive a much larger share of electromagnetic radiation.)} The very first direct detection, indeed, was a {\it short} but impetuous symphony. Because LIGO has not yet reached their designed sensitivity (and because of the --surprisingly high-- total mass of the system) it observed the last tenths of a second prior to merger, or in other words, only a few cycles were seen in the LIGO band before the two black holes blended into one. During the (much shorter) merger phase gravity becomes strong and perturbative methods fail. Hence, once the black holes are nearing collision, numerical simulations are needed. After a breakthrough that occurred around ten years ago, numerical relativity has matured into a very successful area of research. However, it remains technically challenging, since only a relatively small number of orbital cycles (compared to the total number expected within the LIGO band for many astrophysical events) can be modeled within a manageable --weeks to months-- time frame with supercomputers. In fact, large mass ratios and rapid rotation are significantly more difficult to model.  Yet, in many situations of interest, with various mass ratios and large spins, an accurate description of the binary's dynamics is required over many --of the order of thousands-- cycles for earth-based detectors, and many (many) more for future space-based observatories, such as eLISA. Therefore, analytic counterparts remain of vital importance.  Not only to provide a deeper understanding of the dynamics but also to model the entire number of observed cycles. This will allow us to profit the most from the  new era of multi-messenger astronomy, mapping the contents of the universe with unprecedented accuracy. For instance, the knowledge of the distribution of black hole spins could constraint the existence of (very) light particles. Moreover, even though the merger is naturally posed to become a laboratory to test strong gravitational couplings, the analytic control of the inspiral phase --in contrast to the intricate aspects of the~merger-- will also carry a large amount of information. In particular, minute tidal deformations during binary coalescence may leave an imprint in the waveforms which will help us elucidate the inner structure of neutron stars and black holes, and ultimately the dynamics of spacetime.\vskip 4pt
 \begin{wrapfigure}{r}{0.2\textwidth}
\vspace{-0.9cm}
\begin{center}
\includegraphics[width=0.2\textwidth]{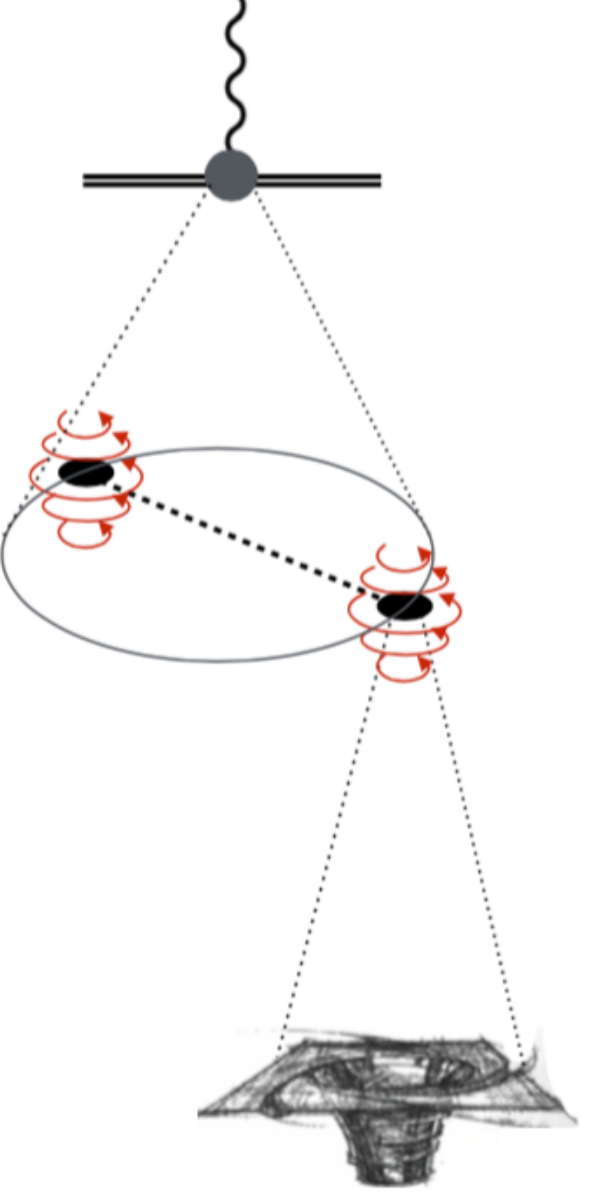}
\end{center}
\vspace{-0.9cm}
\end{wrapfigure}
While perturbative calculations are the bread and butter of theoretical physics, the Post-Newtonian expansion for the two-body problem in general relativity is remarkably complex. Even ignoring spins, and tidal effects (which enter at high orders), the non-linearities in the field equations, and the disparate scales involved, make the enterprise of solving for the motion a real tour de force. In the past, the computations were performed following traditional methods to solve Einstein's equations iteratively. The calculations presented several challenges, on top of tedious algebraic manipulations, such as the appearance of infinities in intermedia steps. The existence of divergences\footnote{Singularities arise from the approximation scheme in which compact bodies (such as black holes) are treated as point-like objects, endowed with a series of `multipole moments', and the long-range (non-linear) nature of gravitational interactions.} required a regularization procedure that, in traditional approaches, also introduced ambiguities due to the use of arbitrary regulators. While many of these issues were eventually resolved (and others remain under discussion) it became apparent that a more systematic framework was desirable, also to be able to incorporate rotational degrees of freedom, which had been largely ignored until recently. Motivated by this, an innovative approach to the two-body problem in gravity was recently developed, bypassing some of the complexity of the traditional methods and providing a more direct connection between the calculations and physical results. 
The novel ingredient is the implementation in gravity of very powerful tools which originated in the study of bound states in Quantum Chromodynamics (QCD), the theory of quarks and gluons describing the strong interaction. The new techniques are collectively referred as the `effective field theory' (EFT) approach, because of the way --deeply rooted in symmetry arguments-- in which  the imprint of physics on short-distance scales is parameterized at long(er) distances.\vskip 4pt 
 \begin{wrapfigure}{r}{0.15\textwidth}
\vspace{-0.9cm}
\begin{center}
\includegraphics[width=0.15\textwidth]{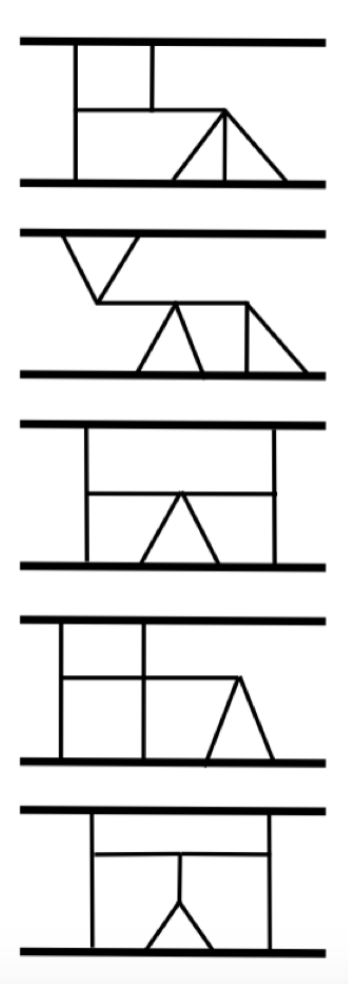}
\end{center}
\vspace{-0.8cm}
\end{wrapfigure}
In many respects gravity and the strong interaction are alike, and bound states of heavy quarks interacting via gluon exchange, and moving non-relativistically, deeply resemble the two-body problem in the Post-Newtonian framework.\footnote{Because of the structure of the strong interaction, bound states do not simply radiate pure glue. They do, instead, radiate photons since quarks have electric charge. However, the radiation problem in gravity is significantly more involved than in electromagnetism. Nevertheless, the EFT techniques can also be extrapolated to the radiation sector.} The main difference is the classical nature of the binary system whereas QCD is drenched on quantum effects. The classical setting, nonetheless, still shares many of the same computational hurdles. It will come then as no surprise that the same techniques which are at the core of computations in particle physics, such as Feynman diagrams, regularization, renormalization, renormalization group flows, as well as more technical developments, have naturally found their way into classical computations in gravitational dynamics within the EFT approach. Calculations using the new framework have reproduced most of the known results for non-rotating binary system in a systematic fashion. (For instance, the first correction to Newtonian's dynamics --responsible for the anomalous precession of Mercury's perihelion-- that took Einstein~and collaborators several pages of calculations, can be readily derived within the EFT approach with only two simple Feynman topologies.) At the same time, the EFT paradigm has been instrumental to~describe spinning compact objects, playing a key~role~extending the state of the art knowledge of the binary's dynamics and emitted power. The calculations for rotating bodies --significantly more challenging--  are much~more recent. For example, the leading spin effects in the binary problem were computed some forty years ago, amusingly following similarities with the spin interaction (Pauli matrices) in the hydrogen~atom. Since the development of the EFT formalism, non-linear spin effects have been incorporated up to high orders, and altogether the state-of-the-art modeling of orbital motion and radiated power for binary systems is approaching a very high level of accuracy (up to order $(v/c)^8$). In principle, this is sufficient for the projected LIGO sensitivity, but for future observatories more precision will be required. The development of analytic methods for the two-body problem in gravity thus remain a thriving and vibrant~field.\vskip 4pt

The LIGO results are among the greatest experimental achievements of all times. Time and again scientists have compared this feat to Galileo pointing his telescope to the sky, offering instead an {\it ear} to the cosmos. After the remarkable landmark of detection, the physics community will soon turn into the study of the properties of the sources, addressing fundamental questions in astrophysics and cosmology. A combined numerical and analytic effort to the binary problem is of paramount importance in light of the nascent program of multi-messenger astronomy. The century of gravitational wave science is in the making --probing the very fabric of spacetime-- and many discoveries are yet to come in the advent of a new era of `precision~gravity'.

\subsection*{Further Reading}

\bit
\im[1.] A.~Zee,  ``Einstein Gravity in a Nutshell,'' Princeton Univ. Pr. (2013).
\im [2.] R.~A.~Porto,  ``The effective field theorist's approach to gravitational dynamics,''  Phys.\ Rept.\  {\bf 633}, 1 (2016). 
\im [3.] R.~A.~Porto,  ``The Tune of Love and the Nature(ness) of Spacetime,'' Fortschritte der Physik {\bf 64}, 10 (2016). 
\eit

\end{document}